Observation of strong bulk damping-like spin-orbit torque in chemically disordered ferromagnetic single layers

Lijun Zhu[1,2]*, Xiyue S. Zhang[1], David A. Muller[1], Daniel C. Ralph[1,2], Robert A. Buhrman[1]
1. Cornell University, Ithaca, New York 14850, USA
2. State Key Laboratory of Superlattices and Microstructures, Institute of Semiconductors, Chinese Academy of Sciences, P.O. Box 912, Beijing 100083, China
3. Kavli Institute at Cornell, Ithaca, New York 14853, USA

Email: lz442@cornell.edu

**Abstract:** Strong damping-like spin-orbit torque ($\tau_{DL}$) has great potential for enabling ultrafast energy-efficient magnetic memories, oscillators, and logic. So far, the reported $\tau_{DL}$ exerted on a thin-film magnet must resut from an externally generated spin current or from an internal non-equilibrium spin polarization in noncentrosymmetric GaMnAs single crystals. Here, we for the first time demonstrate a very strong, unexpected $\tau_{DL}$ from current flow *within* ferromagnetic single layers of chemically disordered, face-centered-cubic CoPt. We establish that the novel $\tau_{DL}$ is a bulk effect, with the strength per unit current density increasing monotonically with the CoPt thickness, and is insensitive to the presence or absence of spin sinks at the CoPt surfaces. This $\tau_{DL}$ most likely arises from a net transverse spin polarization associated with a strong spin Hall effect (SHE), while there is no detectable long-range asymmetry in the material. These results broaden the scope of spin-orbitronics and provide a novel avenue for developing single-layer-based spin-torque memeory, oscillator and logic technologies.

**1. Introduction**
Currently, exerting a nonzero $\tau_{DL}$ on a fFM layer requires an external transverse spin current from the bulk SHE [1-9] of an adjacent spin-orbit layer, an interfacial spin-orbit effect at magnetic interfaces,[8,10-13] or the anomalous Hall effect (AHE)[13] of an adjacent FM layer. Very recently, the generation of SOTs via current flow inside FMs has become an emerging focus of spin orbitronics. The FM layer inside a normal metal (NM)/FM/NM trilayer can experience an *interfacial* $\tau_{DL}$ when the backflow of the planar Hall spin current from opposing interfaces becomes asymmetric due to different spin sinking effectiveness at these interfaces.[14] Non-zero *interfacial* $\tau_{DL}$ on FM layers have been reported when the backflow of the anomalous Hall spin current involves spin rotation at the surfaces.[15] Recently, the so-called "anomalous" SOT effect [16] has been reported to yield a *zero* net $\tau_{DL}$ but finite opposing $\tau_{DL}$ at the surfaces of $Ni_{80}Fe_{20}$ layers that are thicker than its exchange length. So far, a nonzero *bulk* damping-like torque has been reported in strained zinc-blende GaMnAs single crystals with crystal inversion asymmetry [17,18]. However, it remains undetermined as to whether a strong bulk $\tau_{DL}$ can be generated within a single chemically disordered FM layer.

In this work, we for the first time report the presence and the characteristics of strong, unexpected bulk SOTs in single layers of chemically disordered CoPt (= $Co_{50}Pt_{50}$), a FM material with robust spin-orbit coupling (SOC) due to the Pt atoms.[19]

**2. Results and Discussion**
**2.1. Characteristics of strong bulk SOTs**

The CoPt layers with different thicknesses of $t_{CoPt}$ = 2-24 nm are sputter-deposited on oxidized Si substrates. The CoPt layers has in-plane magnetic anisotropy (Section S1, Supporting Information). Using in-plane angle-dependent harmonic response measurement [20,21] under a sinusoidal electric field of $E \approx$16.7 kV/m (see Figure 1a for the measurement geometry), we determine *net* damping-like and field-like SOT efficiencies via

$$\xi^j_{DL(FL)} = (2e/\hbar)\mu_0 H_{DL(FL)} M_s t_{CoPt}/j, \quad (1)$$

where $H_{DL(FL)}$ are damping-like (field-like) SOT effective field (see Section S2, Supporting Information), $\hbar$ is the reduced Plank's constant, $\mu_0$ the permeability, $M_s$ the saturation magnetization of the spin current detector, and $j$ the charge current density of the spin current generator.

As plotted in Figure 1b, when $t_{CoPt}$ exceeds 4 nm a nonzero $\xi^j_{DL}$ emerges and increases monotonically with $t_{CoPt}$. This $\xi^j_{DL}$ behavior is not consistent with it arising from an interfacial SOT since then $\xi^j_{DL}$ would should be approximately independent of $t_{CoPt}$. In Figure 1c, we plot the efficiency of $\tau_{DL}$ per thickness, $\xi^j_{DL}/t_{CoPt}$. This indicates that $\tau_{DL}$ has the characteristic of a bulk effect that has an onset at ~4 nm and then increases in strength until it reaches its full bulk intensity at ~ 14 nm. We have reaffirmed this unique thickness dependence of $\tau_{DL}$ with separate spin-torque ferromagnetic resonance (ST-FMR, Figure 1d) measurements [22]. Figure 1e shows a representative ST-FMR spectrum for the 24 nm CoPt film, which shows a strong symmetric (*S*) component due to $\tau_{DL}$. As discussed in detail in Sections S3 and S4 of the Supporting Information, for a given rf current sourced into a magnetic microstrip, we have

$$\xi^j_{DL}/t_{CoPt} \propto S\sqrt{t_{CoPt}}/C_{MR}\chi, \quad (2)$$

where $C_{MR}$ is the longitudinal magnetoresistance ratio and $\chi$ is a parameter that accounts for the reflection of rf power due to the impdedance mismatch of the device and the rf circuits. The measured values of $S\sqrt{t_{CoPt}}/C_{MR}\chi$ (10 GHz and 15 dBm) for the CoPt films (Figure 1f) exhibit a $t_{CoPt}$ dependence that is well consistent with the $\xi^j_{DL}/t_{CoPt}$ results obtained from the harmonic response measurements (Figure 1c).

We find $\tau_{DL}$ in these CoPt layers is insensitive to the details of the CoPt interfaces, whether those interfaces are formed with neighboring insulating materials ($SiO_2$ and MgO), or with metallic layers that are either spin reflectors (Hf) or spin sinks (Tb and Ru). As summarized in Table 1, $\xi^j_{DL} \approx -0.12$ for all the symmetric stacks MgO/CoPt 16/MgO, Hf/CoPt 16/Hf, Tb/CoPt 16/Tb, and Ru/CoPt 16/Ru and also for the asymmetric $SiO_2$/CoPt 16/MgO. Both the unique thickness dependence and the robust insensitivity to the nature of CoPt interface make the observed $\tau_{DL}$ distinctly different from any interfacial SOTs.[12,15]

The harmonic measurements also determine a field-like torque which, in contrast, is sensitive to the nature of the CoPt interfaces. As plotted in Figure 1b, for a series of $SiO_2$/CoPt /MgO samples, $\xi^j_{FL}$ becomes detectable for $t_{CoPt}$ = 4 nm, and then is essentially constant for all $t_{CoPt}$ values beyond that point. As shown in Table 1, when spin sinks (Tb and Ru) are placed at the two surfaces of the CoFe $\xi^j_{FL}$ is found to be negligible, while it has close to the same value when the interfaces are formed by different spin reflectors (MgO, $SiO_2$, Hf). We interpret this $\xi^j_{FL}$ as being due to the reflection, with some spin rotation, of the spin current that is generated in the bulk of the CoPt due to the bias current.

**2.2. Exceptionally strong spin Hall effect**

The bulk $\xi^j_{DL}$ that we observe requires the generation of an internal spin current (spin accumulation) within the CoPt, and one that has a *net spin polarization* when averaged over the thickness of the material. Spin currents can be generated by either the SHE, the AHE, or the PHE. Here the AHE and the PHE can be ruled out because the in-plane angle-dependent harmonic response technique we employed is sensitive only to transversely polarized spins, as in the SHE, and cannot capture torques due to spin accumulations whose polarization rotates with the external field as in the AHE and PHE (Section S5, Supporting Information).

A strong SHE in the CoPt should of course exert a substantial $\tau_{DL}$ on an adjacent FM layer, as we found to be the case. For multilayer structures that include two FM layers which are both freely rotatable by external fields, the harmonic response analysis is quite challenging due to the entanglement of the Hall voltage responses from the two FMs. Therefore, we used the ST-FMR technique to examine the spin current emission and resultant SOT in multilayers of CoPt 24 /Ti 0.8/FeCoB (FeCoB = $Fe_{0.6}Co_{0.2}B_{0.2}$). The FeCoB, whose thickness was varied from 1.4 to 4 nm, was chosen as the detector of the spin current emitted from the top surface of the CoPt layer because the FeCoB dominates the FMR spectrum of the trilayer stack, while the Ti layer was used to suppress the exchange coupling between the CoPt and the FeCoB layers. If we define the apparent FMR spin-torque efficiency ($\xi_{FMR}$) from the ratio of the symmetric (*S*) and anti-symmetric (*A*) components of the magnetoresistance response of the ST-FMR (Section S3, Supporting Information), the efficiency of the damping-like torque acting on the FeCoB by the external spin current from the CoPt ($\xi_{DL,ext}^{j}$) in the CoPt/Ti/FeCoB sample is determined as the inverse intercept in the linear fit of $1/\xi_{FMR}$ versus $t_{FeCoB}^{-1}$.[23] As shown in Figure 2, we obtained $\xi_{DL,ext}^{j}$ = -0.12 ±0.01 for a CoPt 24/Ti 0.8/FeCoB multilayer, which is twice the amplitude that we measured for a baseline Pt 4/FeCoB bilayer series (≈0.06). We confirmed that the interfaces and thin layers of Ti 0.8/FeCoB bilayer have a negligible torque contribution ($\xi_{DL,ext}^{j}$ ≈0.001, Figure 2) to the overall SOT through ST-FMR on the FeCoB. We note that $\xi_{DL,ext}^{j}$ ≈ -0.12 represents only a lower bound for the internal value of $\theta_{SH}$ for the CoPt because of the cumulative effects of spin attenuation in Ti, interfacial spin backflow [24,25], and spin memory loss [21] on the ST-FMR measurement. The spin memory loss is likely to be substantial as indicated by the significant interfacial magnetic energy density of ≈1.4 erg/cm$^2$ for the Ti/FeCoB interface [21] (≈0.4 erg/cm$^2$ for the Pt/FeCoB interface, Section S6 in the Supporting Information). These results establish that the spin Hall ratio ($\theta_{SH}$) in CoPt is both high and negative in sign. The latter is somewhat surprising since all previous reports of $\theta_{SH}$ for Pt-based non-ferromagnetic HMs,[4,20-23] including PtMn,[7] have given positive values. We speculate that the strong proximity effect, which has been calculated to result in a substantial exchange splitting for Pt, comparable to its SOC energy,[19] is responsible for the reversed sign and enhanced amplitude of $\theta_{SH}$ for the CoPt. Certainly, a theoretical examination of the spin Berry curvature of CoPt would be informative.

The essential question is: what is the mechanism by which the strong SHE in CoPt results in a non-equilibrium net spin population in the bulk of the material and hence in a strong antidamping bulk self-torque within a single CoPt layer? If a magnetic layer has either a mirror symmetry about its midplane or a two-fold rotational symmetry about the current axis, the spin accumulation due to the SHE and its exchange interaction with the magnetization must be equal and opposite on the two sides of the midplane. Any non-zero net bulk spin torque requires that these mirror and rotational symmetries are broken as by a non-centrosymmetric sample structure.

**2.3. Absence of detectable long-range asymmetry**

We have found no evidence for a long-range symmetry breaking in our CoPt films. X-ray diffraction (Figure 3a) and electron diffraction patterns (Figure 3b and Figure S7 in the Supporting Information) reveal that these CoPt layers consist of chemically disordered fcc (A1) polycrystalline grains, which are globally centrosymmetric. There is also no indication of a vertical gradient in grain size, composition, electron scattering, strain, or temperature in the CoPt samples. As revealed by the cross-sectional scanning transmission electron microscopy images in Figure 3b,c, the CoPt layers have good homogeneity across the thickness, without any statistically significant vertical gradient in the dimension of the columnar crystalline grains. There is also no significant vertical composition gradient as indicated by the fairly constant electron energy loss spectrum (EELS) intensities of Co and Pt in the depth profile (Figure 3d) and by the independence of $M_s$ (≈790 emu/cm$^3$) on $t_{CoPt}$ (Figure 3e). Indeed, we find that a strong artificial vertical composition gradient degrades rather than enhances $\tau_{DL}$ in this system. As summarized in Table I, $\xi_{DL}^j$ is 4 times smaller for a control sample of 16 nm Co$_x$Pt$_{1-x}$ layer with $x$ continuously varying from 0.75 to 0.25 ($\xi_{DL}^j$ ≈ -0.031), compared with that of the uniform 16 nm CoPt layer ($x$ = 0.5), and similarly for a 16 nm Co$_x$Pt$_{1-x}$ layer with the reversed composition gradient direction ($\xi_{DL}^j$ ≈ -0.025). This indicates that neither the magnitude nor the sign of the observed $\tau_{DL}$ in the CoPt is caused by a composition gradient. The resistivity ($\rho_{xx}$) of the CoPt layer remains constant at ≈70 μΩ cm as a function of $t_{CoPt}$, until $t_{CoPt}$ is reduced to 2 nm where the interfacial scattering begins to contribute significantly to $\rho_{xx}$.

We do find a significant vertical compressive strain in the CoPt layer as revealed by the fcc (111) lattice plane spacing as measured by x-ray diffraction which is smaller than the bulk value [26] of 2.214 Å (Figure 3g). However, the strain is essentially invariant as a function of $t_{CoPt}$. Averaged electron diffraction patterns from the top and bottom regions of a sufficiently thick CoPt indicate no difference in the strain within the experimental resolution (Section S8, Supporting Information).

Finally, there is no evidence of any significant role of thermal gradient in our samples. $H_{DL}$ for each CoPt layer scales linearly with the input voltage ($V_{in}$) and thus electric field ($E$) even when $V_{in}$ is increased from 1 V to 5 V which results in anomalous Nernst voltage ($V_{ANE}$) and thus the thermal gradient ($\nabla T$) being enhanced by a factor of 23 (Figure 3h,i). This linear behavior clearly excludes any role of the vertical thermal gradient in the generation of the non-equilibrium spin accumulation. In fact, the thermal gradient during the harmonic measurement ($E$ =1.67 kV/cm, $V_{in}$ = 1 V) remains too small (≤ 0.4 mK/nm, Figure 3j and Section S9 in the Supporting Information) to yield a significant variation of the magnetic or structural properties across the film thickness.

Although there is no theory predicting the novel bulk $\tau_{DL}$ reported here, our experimental observations imply that, in magnetic materials with a strong SOC effect (SHE), a strong bulk damping-like torque can be generated without the need of an obvious long-range symmetry breaking. We speculate that the non-equilibrium spin polarization in CoPt could be associated with a "hidden" local inversion asymmetry (e.g. short-range order or local SOC effect) that can cause asymmetric spin accumulation, spin relaxation or exchange interaction. Our findings of the strong SHE (with the spin current independent of its magnetization orientation and transversely polarized) and bulk $\tau_{DL}$ in the metallic ferromagnetic single layers should fundamentally broaden the scope of spin-orbitronics and stimulate future experimental and theoretical efforts on intriguing SOC effects (generation, relaxation,[8] and transport of spin current and local inversion asymmetry) in FMs. The novel bulk $\tau_{DL}$ we demonstrate here may also provide a possible interpretation for the absence of $\tau_{DL}$ in thin $L1_0$-FePt single crystal,[27] the presence of self-induced switching in very thick $L1_0$-FePt single crystal,[28] and the presence of a nonzero $\tau_{DL}$ in Co/Pd multilayer nanowires [29] and in a 15 nm-thick nearly compensated ferrimagnetic Tb-Co film.[30]

## 3. Conclusion

We have demonstrated the first experimental observation of an unexpected, strong $\tau_{DL}$ from current flow *within* ferromagnetic single layers of chemically disordered, face-centered-cubic CoPt. We identify that the novel $\tau_{DL}$ is a bulk effect, with the strength per unit current density increasing monotonically with the FM thickness, and is insensitive to the presence or absence of spin sinks at the CoPt surfaces. This $\tau_{DL}$ most likely arises from a net transverse spin polarization associated with a strong spin Hall effect (SHE), while there is no detectable long-range asymmetry in the material. These results provide a novel avenue for generating strong dampinglike spin-orbit torque in disordered magnetic single layers. This work also establishes the technological possibility of self-driven single-layer spintronics devices (e.g. memory, oscillator, and logic) with simplified architecture and improved scalability.

**4. Experimental Section**

*Sample preparation:* The samples were deposited at room temperature by sputtering onto oxidized silicom substrates with an argon pressure of 2 mTorr and a base pressure of ~$10^{-9}$ Torr. Each sample was capped by a MgO 2 nm/Ta 1.5 nm bilayer that was fully oxidized upon exposure to the atmosphere. The samples were patterned by photolithography and ion milling into 5×60 μm$^2$ Hall bars and 10×20 μm$^2$ microstrips, followed by deposition of 5 nm Ti and 150 nm Pt as electrical contacts for harmonic response measurements and for ST-FMR measurements.

*Structural characterizations*: A Rigaku Smartlab diffractometer was used for x-ray diffraction measurements. A FEI Strata 400 STEM Focused Ion Beam system was used to prepare the sample for scanning transmission electron microscopy (STEM). An Electron Microscope Pixel Array Detector (EMPAD) in FEI F20 STEM (200kV) was used to collect the 4D STEM diffraction data. A FEI/Thermo Fisher Titan Themis STEM system at 300 kV was used for the scanning transmission electron microscopy (STEM) imaging and Electron Energy Loss Spectrum (EELS) measurements.

*Magnetic and electrical measurements*: The saturation magnetization of each sample was measured at 300 K with a vibrating sample magnetometer (sensitivity ~$10^{-7}$ emu) in a Quantum Design physical property measurement system (PPMS). Anomalous Hall voltage and effective anisotropic field were measured electrically using PPMS. During the harmonic response measurements, a Signal Recovery DSP Lock-in Amplifier (Model 7625) was used to source a sinusoidal voltage onto the Hall bars and to detect the first and second harmonic Hall voltage responses. For the spin-torque ferromagnetic resonance measurements, a rf signal generator and a Signal Recovery DSP Lock-in Amplifier (Model 7625) was used and an in-plane magnetic field was swept at 45º with respect to the magnetic microstrip. An electromagnet with maximum in-plane field of 3.5 kOe was used during the spin torque measurements. All the measurements were performed at room temperature.

**Supporting Information**
Supporting Information is available from the Wiley Online Library or from the author.

**Acknowledgements**
The authors gratefully acknowledge P. Kelly, M. D. Stiles, V. P. Amin, Y.-T. Shao for the fruitful discussions. The authors thank R. C. Tapping, Q. Liu, E. Padgett, C. Chang for help with sputtering and STEM measurements, respectively. This work was supported in part by the Office of Naval Research (N00014-15-1-2449), by the NSF MRSEC program (DMR-1719875) through the Cornell Center for Materials Research. This work was performed in part at the Cornell NanoScale Facility, an NNCI member supported by NSF Grant ECCS-1542081.

**Conflict of Interest**
The authors declare no conflict of interest.

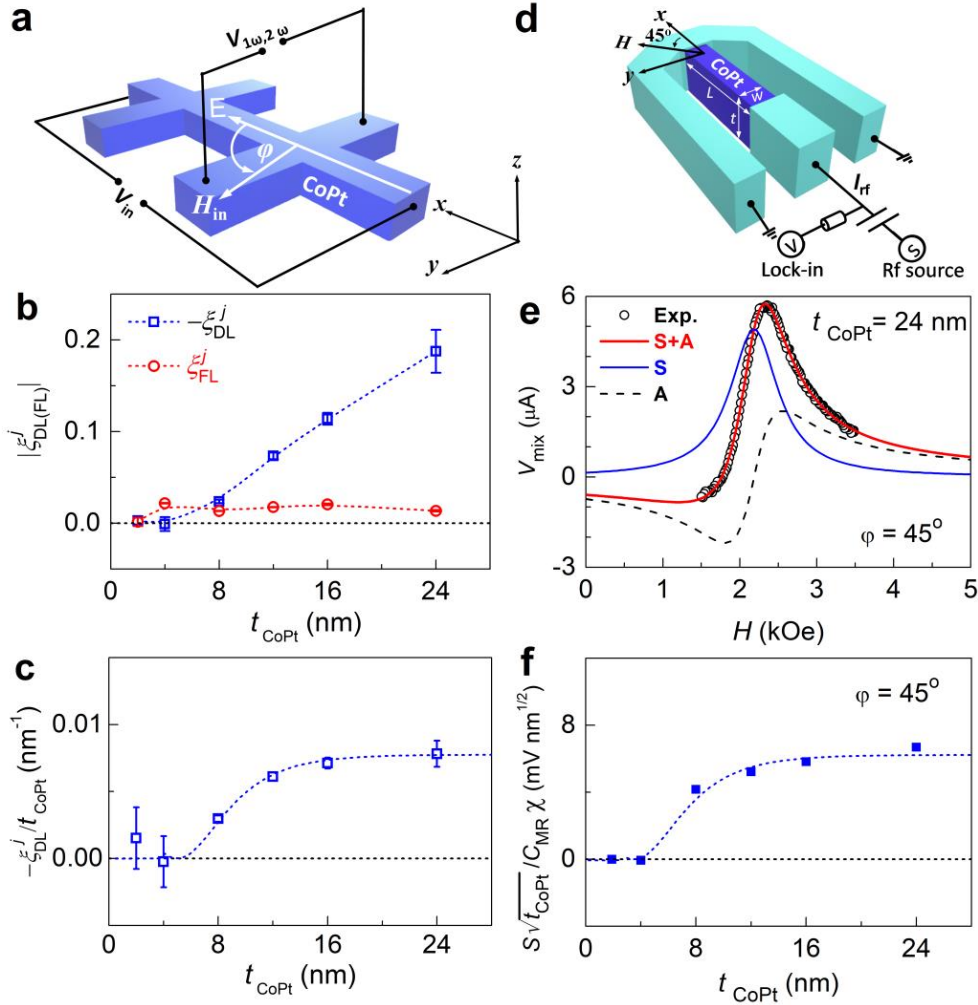

Figure 1. Bulk damping-like spin-orbit torque in CoPt single layers. a) Geometry of "in-plane" harmonic response measurement on the single-layer CoPt. b) Damping-like (field-like) SOT efficiencies per unit current density, $\xi^j_{DL(FL)}$, and c) Damping-like SOT efficiency per thickness ($\xi^j_{DL}/t_{CoPt}$) for different thickness of CoPt single layers. d) Geometry of ST-FMR measurement on the single-layer CoPt. e) ST-FMR spectutra for the 24 nm CoPt single layer, from which a clean symmetric component (solid blue line) that results from damping-like torque can be seen, in addition to the asymmetric component due to the field-like torque. f) $-S\sqrt{t_{CoPt}}/C_{MR}\chi$ (10 GHz, 15 dBm) for the CoPt microstrips plotted as a function of the CoPt thickness ($t_{CoPt}$). In c) the 4 nm CoPt sample has a relatively large error bar, which is mainly due to its high resistivity and small thickness.

**Table 1. Effects of interfaces and vertical composition gradient.** Damping-like (field-like) spin-orbit torque efficiency per unit current density $\xi^j_{DL(FL)}$ as determined from in-plane harmonic response measurements.

| Structure | $\xi^j_{DL}$ | $\xi^j_{FL}$ |
|---|---|---|
| SiO$_2$/CoPt 16/MgO 2 | -0.125 ±0.008 | 0.018±0.002 |
| MgO 2/CoPt 16/MgO 2 | -0.130 ±0.004 | 0.014±0.002 |
| Hf 2.5/CoPt 16/Hf 2.5 | -0.115±0.004 | 0.019±0.002 |
| Tb 2.5/CoPt 16/Tb 2.5 | -0.119 ±0.013 | -0.002±0.002 |
| Ru 1/CoPt 16/Ru 1 | -0.124 ±0.002 | 0.003±0.002 |
| SiO$_2$/Co$_x$Pt$_{1-x}$ ($x$ = 0.75→0.25) 16/MgO | -0.031 ±0.002 | 0.057±0.001 |
| SiO$_2$/Co$_x$Pt$_{1-x}$ ($x$ = 0.25→0.75) 16/MgO | -0.025 ±0.004 | -0.043±0.001 |

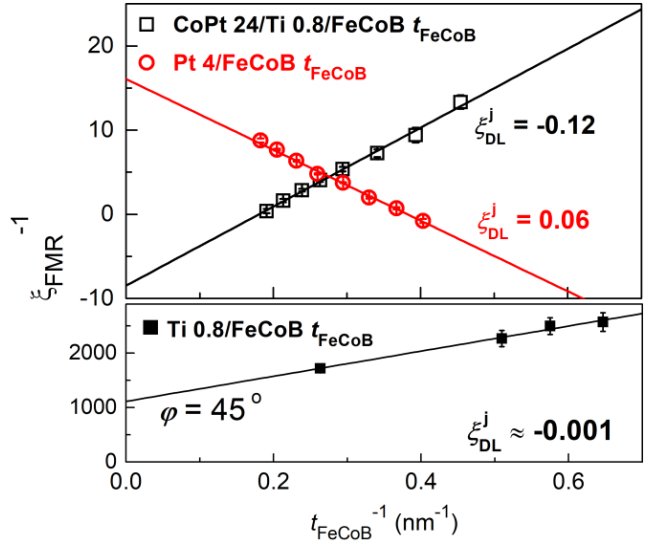

Figure 2. Spin Hall effect in CoPt. Inverse FMR efficiency ($\xi_{FMR}^{-1}$) vs inverse FeCoB thickness ($t_{FeCoB}^{-1}$) for CoPt 24/Ti 0.8/FeCoB $t_{FeCoB}$ and Pt 4/FeCoB $t_{FeCoB}$, and Ti 0.8/FeCoB $t_{FeCoB}$, indicating a very strong spin emission from the CoPt compared with that of Pt.

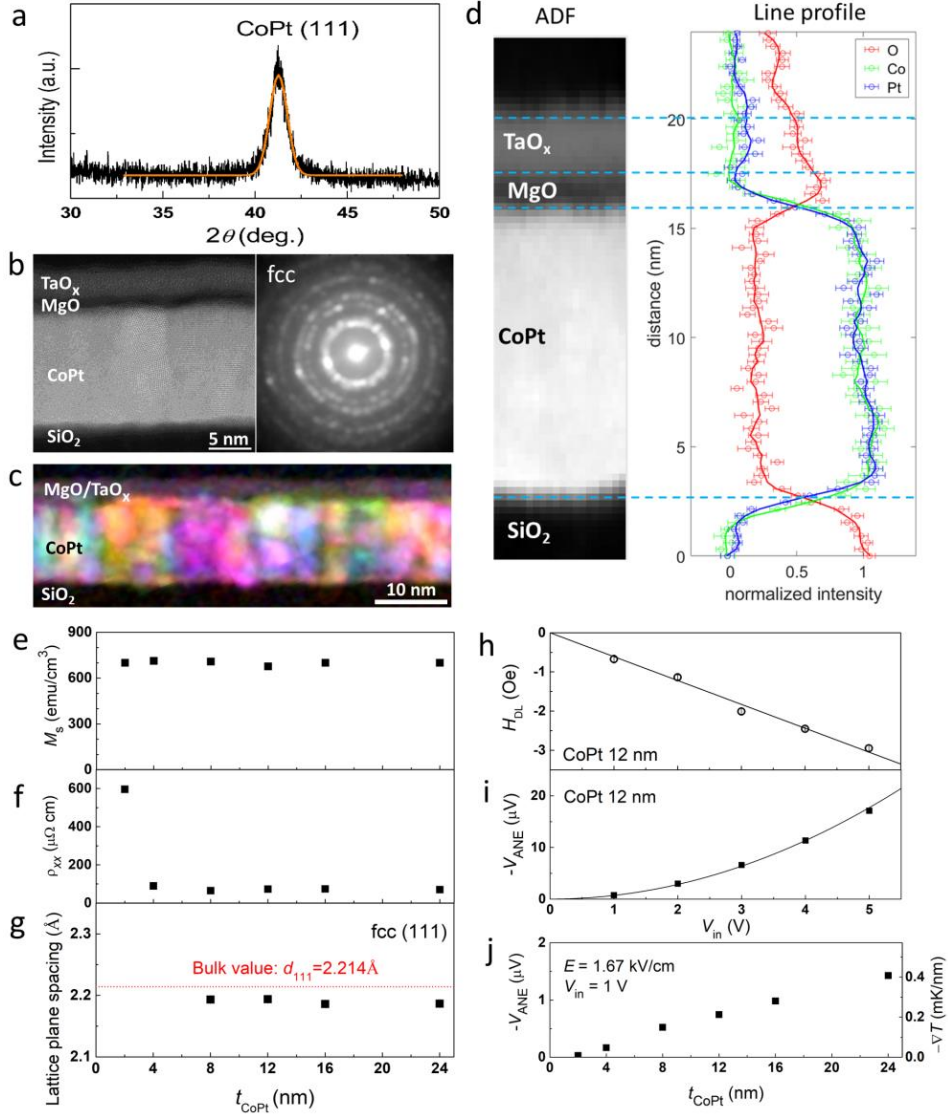

Figure 3. Lack of an apparent inversion asymmetry. a) X-ray diffraction pattern and b) High-angle dark-field cross-sectional STEM image and electron diffraction pattern for a 16 nm CoPt sample, indicating that the CoPt layer has a uniform polycrystalline fcc (111) structure and sharp interfaces with the insulating under- and over-layers. c) Grain distribution with the color showing the different crystalline orientations of the grains, indicating no evidence of vertical shape variation of the columnar grains. d) Depth profile of the EELS intensity for O, Co, and Pt, showing the absence of any composition gradient in the CoPt layer. e) Saturation magnetization, f) resistivity, g) the lattice plane spacing of fcc (111) for the CoPt layers plotted as a function of the CoPt thickness, respectively. The red dashed line in g) represents the bulk value of the lattice plane spacing of 2.214 Å (*27*). h) Damping-like SOT effective field and i) Anomalous Nernst voltage for a 12 nm CoPt layer plotted as a function of input voltage onto a 60 um long Hall bar, indicating the absence of any effects of the vertical thermal gradient on the damping-like spin torque. j) The CoPt thickness dependence of the anomalous Nernst voltage and the thermal gradient in the CoPt layer under an electric field of $E$ =1.67 kV/cm ($V_{in}$ = 1 V).

9